\documentclass[a4paper, oneside]{article}
\usepackage[utf8]{inputenc}  
\usepackage[english]{babel}
\usepackage[T1]{fontenc} 
\usepackage{lmodern}
\usepackage[tbtags]{amsmath}
\usepackage{amsfonts,amssymb,mathrsfs,amscd}
\usepackage{amsthm}
\usepackage{graphicx}
\usepackage{cite}
\usepackage{hyperref}
\hypersetup{
	colorlinks=true,
	linkcolor=blue,
	filecolor=blue,
	urlcolor=blue,
} 

\newcommand{\Tr}{{\rm Tr}}
\newtheorem{definition}{Definition} 
\newtheorem{statement}{Statement}
\newtheorem{algorithm}{Algorithm}   
 
\begin{document}
	
\title{\bf Numerical estimation of reachable \\
and controllability sets \\
for a two-level open quantum system \\
driven by coherent and incoherent controls\footnote{arXiv version of the article [Oleg V. Morzhin and Alexander N. Pechen, 
	``Numerical estimation of reachable and controllability sets for a two-level open quantum 
	system driven by coherent and incoherent controls'',  AIP Conference Proceedings, 
	2362, 060003 (2021), \url{https://doi.org/10.1063/5.0055004} ].  This article may be downloaded 
	for personal use only. Any other use requires prior permission of the authors and AIP Publishing.} }

\author{\normalsize {\bf \large Oleg~V.~Morzhin}\footnote{\url{http://www.mathnet.ru/eng/person30382}}~$^{,1}$ \quad and \quad 
	{\bf \large Alexander~N.~Pechen}\footnote{\url{http://www.mathnet.ru/eng/person17991} ; \quad e-mail: {\tt apechen@gmail.com} (corresponding author).}~$^{,1,2}$ \vspace{0.2cm} \\
	\small $^1$ Department of Mathematical 
		Methods for Quantum Technologies,\\
	\small Steklov Mathematical Institute of Russian Academy of Sciences, \\
	\small 8 Gubkina Str., Moscow, 119991, Russia, \\
	\small \url{http://www.mi-ras.ru/eng/dep51}; \\
	\small $^2$ National University of Science and Technology ``MISiS'',\\
	\small 6 Leninskiy prospekt, Moscow, 119991, Russia} 

\date{}	

\maketitle
     
\begin{abstract}
The article considers a two-level open quantum system,  
whose evolution is governed by the Gorini--Kossakowski--Lindblad--Sudarshan master 
equation with Hamiltonian and dissipation superoperator depending, correspondingly,
on piecewise constant coherent and incoherent controls with constrained magnitudes. 
Additional constraints on controls' variations are also considered. 
The  system is analyzed using Bloch parametrization of the system's density matrix. We adapt
the section method for obtaining outer parallelepipedal and pointwise estimations 
of reachable and controllability sets in the Bloch ball via solving a number of problems for optimizing coherent and incoherent controls with respect to some objective criteria. The differential evolution and dual annealing optimization methods
are used. The numerical results show how the reachable sets' estimations depend on 
distances between the system's initial states and the Bloch ball's center 
point, 
final times, constraints on controls' magnitudes and variations. 
\end{abstract}

\tableofcontents

\section{\label{section1}Introduction}

Quantum control, i.e. control of individual quantum objects (atoms, molecules, etc.) 
attracts nowadays high interest both for fundamental reasons
and due to multiple existing and prospective 
applications in quantum technologies~\cite{Brif_Chakrabarti_Rabitz_article_2010, Moore2011, Glaser_Boscain_Calarco_et_al_2015, CPKoch_2016_OpenQS, Morzhin_Pechen_Review_2019}. 
Quantum control theory considers  quantum systems governed by
Schr\"{o}dinger, Liouville--von Neumann, 
Gorini--Kossakowski--Lindblad--Sudarshan, and other
quantum-mechanical equations with controls, and exploits various results 
from the general optimal control theory. For example, necessary and 
sufficient conditions for pure-state/ equivalent-state 
controllability for multilevel quantum systems whose dynamics is described 
by the Schr\"{o}dinger equation with the Hamiltonian linearly depending on coherent control
function, were expressed in terms of special unitary and symplectic Lie algebras~\cite{Albertini_DAlessandro_2003}. Often in real situations controlled quantum systems are open, i.e. interacting with the environment. Important results about controllability of open quantum systems were also obtained, including detailed investigation of controllability for Markovian open quantum systems subject to coherent control~\cite{Altafini2003-1, Altafini2003-2}, construction of universally optimal Kraus maps~\cite{Wu2007} and proving approximate controllability of generic open quantum systems driven by coherent and incoherent controls~\cite{Pechen2011}. Typical 
optimal control problems (OCPs) for quantum systems  include transferring 
an initial quantum state to a given target quantum state, maximizing mean 
value of a quantum observable, 
generating unitary gates, maximizing overlap between system's density matrix and a given target density matrix.

For open quantum systems, there are two general types of control actions. 
Coherent control is typically realized by laser radiation. Incoherent control is realized, e.g., using state of incoherent environment in the dissipative part
of the master equation~(\cite{Pechen_Rabitz_2006} and \cite{Pechen2011, Pechen_Rabitz_2014}), back-action of non-selective quantum measurements~\cite{Pechen_Ilin_Shuang_Rabitz_2006}, combining quantum measurements
and quantum reinforcement learning~\cite{Dong_Chen_Tarn_Pechen_Rabitz_2008}, in 
purely dissipative dynamical equation (i.e. without non-dissipative term in the right-hand side of the Gorini--Kossakowski--Lindblad--Sudarshan master equation) with controlled dissipator~\cite{Basilewitsch_Koch_Reich_2019}.

The papers~\cite{Pechen_Rabitz_2006} and \cite{Pechen2011, Pechen_Rabitz_2014} proposed and developed the general method of incoherent control of open quantum systems via engineered environment, which can be used independently or together with coherent control. In the article~\cite{Pechen2011}, this 
approach was applied for developing a method
for realizing approximate controllability of open quantum systems in the set of all density matrices. Based on the articles~\cite{Pechen_Rabitz_2006} and \cite{Pechen2011, Pechen_Rabitz_2014},
a two-level open quantum system driven by coherent $v$ and incoherent $n$ 
controls was written in our article~\cite{Morzhin_Pechen_IJTP_2019}. For the corresponding time-minimal control problem, the paper~\cite{Morzhin_Pechen_IJTP_2019} describes the approach based on reducing this OCP to a series of auxiliary OCPs, each of them is defined for a unique final time from a series 
$\{T_i\}$, with the objective functional being square of the 
Hilbert--Schmidt distance between the final density matrix and a given target density matrix. For an auxiliary OCP, it was suggested to use the two-parameter gradient projection method,
which is long time known in the general optimal control theory~\cite{Demyanov_Rubinov_book_1970}. Articles~\cite{Morzhin_Pechen_LJM_2019, Morzhin_Pechen_Physics_of_Particles_and_Nuclei, 
	Morzhin_Pechen_LJM_2020, Morzhin_Pechen_SteklovProceedings} considered for the same 
two-level open quantum system time-minimal control, use of 
different optimization methods, checking different conditions of optimality of controls, 
generation of suboptimal final times and controls via machine learning, analytical 
exact description of reachable sets, etc. 

An important problem in quantum control is to describe (exactly or approximately) 
reachable sets (RSs) and controllability sets (CSs) for a controlled system in  the spaces of pure or mixed quantum states~\cite{DAlessandro2000, 
Boussaid_Caponigro_Chambrion_2012, Yuan_Automatica_2013, Li_Lu_Luo_Laflamme_Peng_Du_2016, ShackerleyBennett_Pitchford_Genoni_Serafini_Burgarth_2017, Dirr_vom_Ende_Schulte-Herbruggen_2019}. For the open two-level quantum system, which was considered in~\cite{Morzhin_Pechen_IJTP_2019, Morzhin_Pechen_LJM_2019, Morzhin_Pechen_Physics_of_Particles_and_Nuclei, Morzhin_Pechen_LJM_2020, 
Morzhin_Pechen_SteklovProceedings}, this article analyzes its RSs and CSs 
in the terms of the Bloch parametrization, i.e. via RSs and CSs of the corresponding dynamical system whose states are Bloch vectors. Because these vectors  are located in the unit ball $\mathcal{B} := \{ x \in \mathbb{R}^3: \| x\|_2^2 \leq 1\}$, the problem of estimating RSs and CSs in the space of density matrices is reduced to
the simpler problem of estimating RSs and CSs for the derived system. For 
solving the latter problem, we adapt the section method (see~\cite{Morzhin_Tyatyushkin_2008, Tyatyushkin_Morzhin_2011}), which is based on solving 
a series of OCPs. We consider piecewise constant controls $v,n$ that means that the objective 
becomes  function of the corresponding finite-dimensional vector argument. For minimization of the objective function, two stochastic zeroth-order optimization methods have been used, 
differential evolution method (DEM)~\cite{scipy_differential_evolution, Storn_Price_1997} 
and dual annealing method (DAM)~\cite{scipy_dual_annealing, Tsallis_Stariolo_1996, Xiang_Gong_2000} both known in the theory of global optimization. 

The structure of the article is the following. In Section~\ref{section2}, 
the quantum system and various types of constraints on controls are formulated. The dynamical system whose states are Bloch vectors is written in Section~\ref{section3}. Section~\ref{section4} formulates the definitions 
of RSs and CSs taking into account the additional constraints on controls. Section~\ref{section5} defines outer parallelepipedal and pointwise estimations of RSs and CSs, formulates two estimating algorithms.  Section~\ref{section6} is devoted to using DEM and DAM. Section~\ref{section7} describes our numerical results. The Conclusions section~\ref{Conclusions} 
resumes the article.

\section{Quantum System. Constraints on Controls}\label{section2}

The articles~\cite{Pechen_Rabitz_2006} and \cite{Pechen2011, Pechen_Rabitz_2014}
consider multi-level quantum systems with coherent and incoherent controls 
and arbitrary number of levels, at that such a two-level model as an example was analyzed
in~\cite{Pechen2011} (calcium atom). Based on these articles, 
the work~\cite{Morzhin_Pechen_IJTP_2019} considers the following two-level model,
which afterwards was analyzed also in our papers~\cite{Morzhin_Pechen_LJM_2019,
Morzhin_Pechen_Physics_of_Particles_and_Nuclei, Morzhin_Pechen_LJM_2020, 
Morzhin_Pechen_SteklovProceedings}. Consider the Gorini--Kossakowski--Lindblad--Sudarshan master equation 
\begin{eqnarray} 
	\frac{d \rho(t)}{dt} &=&
	-\frac{i}{\hbar} \Big[ \widehat{\bf H}_{v(t)}, \rho(t) \Big] + 
	\mathcal{L}_{n(t)}(\rho(t)), 
	\qquad \rho(t) \in \mathbb{C}^{2 \times 2}, \qquad \rho(0) = \rho_0. 
	\label{f_1}
\end{eqnarray}
Here $\rho(t)$ is the density matrix, i.e. a Hermitian positive semi-definite,
$\rho(t) = \rho^{\dagger}(t) \geq 0$, with unit trace, ${\rm Tr}\rho(t) 
= 1$.
The Hamiltonian linearly depends on coherent control $v$: 
\begin{eqnarray}
	\widehat{\bf H}_{v(t)} = \widehat{\bf H}_0 + \widehat{\bf H}_1 v(t), \qquad 
	\widehat{\bf H}_0, ~\widehat{\bf H}_1 \in \mathbb{C}^{2 \times 2}; 
	\label{f_2}
\end{eqnarray} 
the controlled dissipative superoperator acts on the density matrix as
\begin{eqnarray} 
	\mathcal{L}_{n(t)} (\rho(t)) &=& 
	\gamma \left( n(t) + 1 \right) \left( \sigma^- \rho(t) \sigma^+ - \dfrac{1}{2} \left\{ \sigma^+ \sigma^-, \rho(t) \right\} \right) + \nonumber \\
	&& + \gamma n(t) \left( \sigma^+ \rho(t) \sigma^- - \dfrac{1}{2} \left\{ 
\sigma^- \sigma^+, \rho(t) \right\} \right), \qquad \gamma > 0. 
	\label{f_3}
\end{eqnarray}
The free Hamiltonian $\widehat{\bf H}_0$ is assumed to have different eigenvalues. 
The Hamiltonian $\widehat{\bf H}_1$ describes the interactions 
between coherent control and the quantum system;
$\mathcal{L}_{n}(\rho)$ describes the controlled interactions 
between the quantum system and its environment (reservoir). Matrices $\sigma^+ = \begin{pmatrix}
0 & 0 \\ 1 & 0
\end{pmatrix}$, $\sigma^- = \begin{pmatrix}
0 & 1 \\ 0 & 0
\end{pmatrix}$ define the transitions between the two energy levels of the quantum system, $n$ is the incoherent control. 
The notations $[A,B] = AB - BA$  and $\{A,B\} = AB + BA$ mean, correspondingly, the commutator and anti-commutator  of two operators $A,B$. Without loss of generality one can consider the free Hamiltonian 
$\widehat{\bf H}_0 = \hbar \omega \begin{pmatrix}
0 & 0 \\
0 & 1
\end{pmatrix}$ and interaction Hamiltonian 
$\widehat{\bf H}_1 = \mu \sigma_1$, where $\omega > 0$, $\mu \in \mathbb{R}$, $\mu \neq 0$;  
$\hbar$ is the Planck's constant; $\sigma_1 = 
\begin{pmatrix}
0 & 1 \\ 1 & 0
\end{pmatrix}$ is one of the Pauli matrices, other Pauli matrices are
$\sigma_2 = \begin{pmatrix}
0 & -i \\ i & 0
\end{pmatrix}$,
$\sigma_3 = \begin{pmatrix}
1 & 0 \\ 0 & -1
\end{pmatrix}$. 

The initial density matrix $\rho_0$ is fixed for the problem of describing the system's RSs. The initial density matrix is not fixed for the problem of describing the system's CSs; this case requires fixing target density~matrix. 

Coherent $v$ and incoherent $n$ controls are considered as scalar functions. They form vector control $u=(v,n)$ which satisfies the pointwise constraint 
\begin{eqnarray} 
	u(t) = (v(t), n(t)) \in Q := \big[v_{\min}, v_{\max}\big] \times \big[0, n_{\max}\big]
	\label{f_4}
\end{eqnarray} 
at the whole time range $[0,T]$, where the bounds $v_{\min}$, $v_{\max}$, 
$n_{\max}$ are given.

By analogy with works~\cite{Morzhin_Pechen_LJM_2019, Morzhin_Pechen_Physics_of_Particles_and_Nuclei, Morzhin_Pechen_LJM_2020}, in this article we consider piecewise constant controls 
\begin{eqnarray}
	v(t) &=& \sum\limits_{j=0}^{N_v-1} \chi_{[t_j, t_{j+1})}(t_j^v) v_j, 
\quad t \in [0, T), 
	\qquad v(T) = v(T-), 
	\label{f_5} \\
	n(t) &=& \sum\limits_{j=0}^{N_n-1} \chi_{[t_j, t_{j+1})}(t_j^n) n_j, 
	\quad t \in [0, T), 
	\qquad n(T)=n(T-),
	\label{f_6} 
\end{eqnarray}
i.e. for each control one has a uniform distribution of time nodes, $t_j^{v|n} = j (\Delta t)^{v|n}$, $(\Delta t)^{v|n} = T/N_{v|n}$, $j = \overline{0, N_{v|n}}$. These nodes correspond to some given final 
time $T>0$ and natural numbers $N_v, N_n$. 

For representing control $u$ in terms of finite-dimensional optimization, consider the vector
\begin{eqnarray}
	{\bf u} := \left( \left\{ v_j \right\}_{j=0}^{N_v-1}, 
	\left\{ n_j \right\}_{j=0}^{N_n-1} \right) \in Q(N_v,N_n) := [v_{\min}, v_{\max}]^{N_v} \times [0, n_{\max}]^{N_n}, 
	\label{f_7}
\end{eqnarray} 
where $Q(N_v,N_n)$ is a  $(N_v+N_n)$-dimensional compact search space in $\mathbb{R}^{N_v+N_n}$. 

From physical point of view, it can be useful to constrain variations of piecewise constant controls. The constraints on controls' magnitudes (see~(\ref{f_4})) also restrict controls' variations.
Moreover, by analogy with our
papers~\cite{Morzhin_Pechen_Physics_of_Particles_and_Nuclei, Morzhin_Pechen_LJM_2020},
this article considers the following additional constraints on controls' variations.

The first type of additional constraints on controls' variations requires 
to add the regularizer
\begin{eqnarray}
	\mathcal{R}^{\rm Var}(u; \beta^{\rm Var}_{dv}, \beta^{\rm Var}_{dn}) := 
	\beta^{\rm Var}_{dv} {\rm Var}_{[0,T]}(v; N_v)   
	+ \beta^{\rm Var}_{dn} {\rm Var}_{[0,T]}(n; N_n)  
	\label{f_8}
\end{eqnarray}	
to an objective functional to be minimized, where the weight coefficients 

$\beta^{\rm Var}_{dv}, \beta^{\rm Var}_{dn} > 0$; variations of controls~$v,n$ are 
\begin{eqnarray}
	{\rm Var}_{[0,T]}(v; N_v) := \sum\limits_{j=1}^{N_v-1} |v_j - v_{j-1}|, 
	\qquad 
	{\rm Var}_{[0,T]}(n; N_n) := \sum\limits_{j=1}^{N_n-1} |n_j - n_{j-1}|.
	\label{f_9}
\end{eqnarray} 

The second type of additional constraints on controls'  variations means to add the regularizer
\begin{eqnarray}
	\mathcal{R}^{\rm abs}(u; \beta^{\rm abs}_{v}, \beta^{\rm abs}_{n}) := 
	\beta^{\rm abs}_v \sum\limits_{j=0}^{N_v-1} |v_j| + 
	\beta^{\rm abs}_n \sum\limits_{j=0}^{N_n-1} n_j  
	\label{f_10}
\end{eqnarray}
to an objective functional to be minimized, where the weight coefficients 
$\beta^{\rm abs}_v, \beta^{\rm abs}_n > 0$, and we use that $|n_j| = n_j$ in the frames of compact~$Q(N_v,N_n)$. 

The third type of additional constraints requires to satisfy the inequalities
\begin{eqnarray} 
	|v_j - v_{j-1}| \leq \delta_{dv}, \quad 1 \leq j \leq N_v-1, \quad  |n_j 
- n_{j-1}| \leq \delta_{dn}, \quad 1 \leq j \leq N_n-1,
	\label{f_11}
\end{eqnarray}
with some given thresholds $\delta_{dv} \in (0, v_{\max}-v_{\min})$, $\delta_{dn} \in (0, n_{\max})$.
For taking into account these constraints, we form the values
\begin{eqnarray}
	&&M^{dv} := \max\limits_{1 \leq j \leq N_v-1} \left\{ |v_j - v_{j-1}| \right\} \geq 0, \qquad 
	M^{\delta_{dv}} :=  \max \{ M^{dv} - \delta_{dv}, 0 \} \geq 0,  
	\label{f_12} \\
	&&M^{dn} := \max\limits_{1 \leq j \leq N_n-1} \left\{ |n_j - n_{j-1}| \right\} \geq 0, \qquad 
	M^{\delta_{dn}} := \max \{ M^{dn} - \delta_{dn}, 0 \} \geq 0
	\label{f_13}
\end{eqnarray}
and the regularizer
\begin{eqnarray}
	\mathcal{R}^{\max}(u; \beta_{dv}^{\max}, \beta_{dn}^{\max}) := 
	\beta_{dv}^{\max} M^{\delta_{dv}} + \beta_{dn}^{\max} M^{\delta_{dn}},
	\label{f_14}
\end{eqnarray}
where the weight coefficients $\beta_{dv}^{\max}, \beta_{dn}^{\max} > 0$. 
If all the inequalities for $v_j$ and $n_j$ in (\ref{f_11}) are satisfied, then it means, correspondingly, $M^{\delta_{dv}} = 0$ and $M^{\delta_{dn}} = 0$. 

The values $v_{\min}$, $v_{\max}$, $n_{\max}$, $N_v$, $N_n$, $\beta^{\rm Var}_{dv}$, $\beta^{\rm Var}_{dn}$, $\beta^{\rm abs}_v$, $\beta^{\rm abs}_n$, $\beta_{dv}^{\max}$, $\beta_{dn}^{\max}$, $\delta_{dv}$, $\delta_{dn}$ 
allow to define some certain class of admissible controls $v, n$, at that 
$N_v, N_n$ allows to regulate the dimension of the search space~$Q(N_v,N_n)$.  

In a part of the article~\cite{Morzhin_Pechen_Physics_of_Particles_and_Nuclei},
we considered piecewise constant controls~$v,n$ with regularization of the type~(\ref{f_8}) in the composite objective function  to be minimized [formulas~(17), (18) in the work~\cite{Morzhin_Pechen_Physics_of_Particles_and_Nuclei}], which takes into account the goals to minimize both the Uhlmann–Jozsa fidelity and non-fixed final time~$T$. In our article~\cite{Morzhin_Pechen_LJM_2020}, regularizers of the both types~(\ref{f_10}) and (\ref{f_14}) were used in the composite objective function to be minimized [formula~(18) in the work~\cite{Morzhin_Pechen_LJM_2020}], which takes into account also several minimization goals.  

In the next sections of this article, the constraints~(\ref{f_4})--(\ref{f_14}) are used. For further considerations, it is convenient to use the abstract notation $\mathcal{U}([0,T],Q)$ meaning some class of controls, at least, with only~(\ref{f_4}). In the next sections, we mention the certain meaning of $\mathcal{U}([0,T],Q)$, when it is needed. Further we consider only the case $N_v = N_n$.

\section{Dynamics Using the Bloch Parametrization}\label{section3}

For a density matrix $\rho \in \mathbb{C}^{2 \times 2}$,
consider its Bloch parametrization (e.g., \cite{Holevo_book_De_Gruyter_2019})
\begin{eqnarray}
	\rho = \frac{1}{2} \left( \sigma_0 + 
	\sum\limits_{j=1}^3 x_j \sigma_j \right) =
	\frac{1}{2} \begin{pmatrix}
		1 + x_3 & x_1 - i x_2 \\ x_1 + i x_2 & 1 - x_3
	\end{pmatrix},
	\label{f_15}
\end{eqnarray}
where the matrices $\sigma_0 = \mathbb{I}_2$, $\sigma_j$, $j=1,2,3$, form the Pauli basis; Bloch vector 
$x = (x_1, x_2, x_3) \in \mathcal{B}$, 
$x_j = \Tr\left(\rho \sigma_j \right)$, 
$j=1,2,3$. Using the Bloch parametrization, the following dynamical system  
corresponding to the initial system (\ref{f_1}) was obtained in the article~\cite{Morzhin_Pechen_IJTP_2019}:
\begin{eqnarray} 
	\frac{dx(t)}{dt} = \left(A + B^v v(t) + B^n n(t) \right)x(t) + d, 
	\quad x(0) = x_0 \in \mathcal{B}, 
	\label{f_16}
\end{eqnarray}
where, for the given above matrices ${\bf H}_0$, ${\bf H}_1$, we have
\begin{eqnarray} 
	A = \begin{pmatrix}
		-\frac{\gamma}{2} & \omega & 0 \\
		-\omega & -\frac{\gamma}{2} & 0 \\
		0 & 0 & -\gamma
	\end{pmatrix}, \quad 
	B^v = 
	\begin{pmatrix}
		0 & 0 & 0 \\
		0 & 0 & -2 \kappa \\
		0 & 2\kappa & 0
	\end{pmatrix}, \quad 
	B^n = 
	\begin{pmatrix}
		-\gamma & 0 & 0 \\
		0 & -\gamma & 0 \\
		0 & 0 & -2\gamma
	\end{pmatrix}, \quad
	d = \begin{pmatrix}
		0\\
		
		0\\
		\gamma
	\end{pmatrix},  
	\label{f_17} 
\end{eqnarray}
and $\kappa := \mu/\hbar$. The Bloch parametrization (\ref{f_15}) gives the bijection between matrix $\rho(t)$ and the corresponding state $x(t)$ of the system~(\ref{f_16}), and vice versa. If $\|x(t)\|_2 = 
1$, then the corresponding density matrix  $\rho(t)$ describes a pure quantum state, while for $\| x(t) \|_2 < 1$ the corresponding density matrix 
represents a mixed quantum state. The center point $x_O=(0,0,0)$ represents the completely mixed quantum state with density matrix $\rho_O = \mathbb{I}_2/2$, which has entropy $S(\rho_O) = -{\rm Tr}(\rho_O \log_2 \rho_O) = 1$. In contrast, the north pole point $x_N = (0,0,1)$ corresponds to the density matrix $\rho_N = \begin{pmatrix}
1 & 0 \\
0 & 0
\end{pmatrix}$, which has entropy $S(\rho_N) = -{\rm Tr}(\rho_N \log_2 \rho_N) = 0$.
The system~(\ref{f_16}) was considered also in our articles~\cite{Morzhin_Pechen_LJM_2019, Morzhin_Pechen_Physics_of_Particles_and_Nuclei, Morzhin_Pechen_LJM_2020}.

\section{Definitions of Reachable and Controllability Sets} \label{section4}

The function $f(x,v,n) := \left(A + B^v v + B^n n \right)x + d$, which defines the right-hand side in~(\ref{f_16}), is continuous in its arguments. Taking values of some piecewise constant   controls $v$, $n$ instead of the variables $v,n$ over $[0, T]$, we have the function $f(x,t) := f(x,v(t),n(t))$ being continuous in~$x$ and discontinuous in~$t$. This fact violates a key assumption of the classical theorem on existence and uniqueness of Cauchy problems' solutions. For the system~(\ref{f_16}), its solution for some bounded controls $v,n$ is considered in the more general 
meaning based on the Carath\'{e}odory's theorem in the theory of differential equations with discontinuous right-hand sides~\cite{FilippovAF_book_1988}.   

We define RS and CS in the terms of the derived system~(\ref{f_16}). 
It is easy to analyze RSs of the system~(\ref{f_16}) than RSs of the initial system~(\ref{f_1}), because in the former case these sets are in the Bloch ball.  Since in addition to the constraint (\ref{f_4}) we consider the regularizers (\ref{f_8}), (\ref{f_10}), and (\ref{f_14}), then the following definitions of RSs and CSs take into account these regularizers. Thus, the definitions of RSs and CSs differ from usual definitions~\cite{Morzhin_Pechen_SteklovProceedings}.

Define the function
\begin{eqnarray}
	M(x, \widehat{x}; \delta_{x_T}) :=
	\left( \max\left\{ \| x - \widehat{x} \|_p^p - \left( \delta_{x_T} \right)^p, 0 \right\} \right)^p \leq 2^p - \left(\delta_{x_T} \right)^p < 2^p, 
\qquad p \in \{1,2\},
	\label{f_18}
\end{eqnarray}
where $x, \widehat{x} \in \mathcal{B}$; $\delta_{x_T}  \geq 0$, at that $M(x, \widehat{x}; \delta_{x_T} = 0) := \| x - \widehat{x} \|_p^p$; the norm $\| x\|_p = \left(\sum_{j=1}^n \left|x_j \right|^p \right)^{1/p}$.

\begin{definition}[RSs]
	\label{definition1}
	If the system (\ref{f_16}) evolving over a certain range $[0, T]$ is considered with controls which satisfy only the constraint~(\ref{f_4}), then 
RS $\mathcal{R}(T, x_0, \mathcal{U}([0,T],Q))$ at $t = T$ is defined as 
the set of final states $\{x(T) := x(T|~u)\}$ obtained by solving the 
	system~(\ref{f_16}) with a given initial state~$x_0$ for all  
	admissible controls, $u \in \mathcal{U}([0,T],Q)$, i.e. 
	$\mathcal{R}(T, x_0, \mathcal{U}([0,T],Q)) := 
	\bigcup\limits_{u \in \mathcal{U}([0,T],Q)} x(T~|~u)$ in~$\mathcal{B}$, 
	where $x(\cdot |~u)$ denotes the system's solution for a given~$u$. If we consider the class 
	$\mathcal{U}([0,T],Q)$ that consists of controls satisfying~(\ref{f_4}) and~(\ref{f_11}), then RS is formed by all such points, $\{ \widetilde{x} 
\}$, that for each of them there exists such control process $(x(\cdot),u(\cdot))$ that the system of equalities
	\begin{eqnarray}
		M(x(T|u), \widetilde{x}; \delta_{x_T}) = 0, \quad M^{dv}=0, \quad M^{dn} = 0,
		\label{f_19}
	\end{eqnarray}
	is satisfied, where $\delta_{x_T}=0$ is taken. If the regularizer (\ref{f_8}) or (\ref{f_10}) is considered, then RS is defined by the following. For any point $\widetilde{x}$ belonging to the RS there is such control $u=\widetilde{u}$, which simultaneously satisfies~(\ref{f_4}) and
	solves, correspondingly, the minimization problem		
	\begin{eqnarray}
		\mathcal{R}^{\rm Var}(u; \beta^{\rm Var}_{dv}, \beta^{\rm Var}_{dn})
		\to \min\limits_{u}, \quad \text{s.t.} \quad M(x(T|u), \widetilde{x}; \delta_{x_T}) = 0
		\label{f_20}	
	\end{eqnarray}
	or the minimization problem 
	\begin{eqnarray}  
		\mathcal{R}^{\rm abs}(u; \beta^{\rm abs}_{dv}, \beta^{\rm abs}_{dn})
		\to \min\limits_{u}, \quad \text{s.t.} \quad M(x(T|u), \widetilde{x}; \delta_{x_T}) = 0, 
		\label{f_21}	
	\end{eqnarray}
	where $\delta_{x_T} = 0$ is considered. For the problems~(\ref{f_20}), 
(\ref{f_21}), control $u$ is considered in the class of controls~(\ref{f_5}), (\ref{f_6}) satisfying only the constraint~(\ref{f_4}). 
\end{definition} 

For the minimization problems~(\ref{f_19})--(\ref{f_21}), it is suggested, correspondingly, that the following composite objective functionals to be minimized:
\begin{eqnarray}
	\Phi^{\max}(u; \beta^{\max}_{x_T}, \beta^{\max}_{dv}, \beta^{\max}_{dn}) 
&:=& 
	\beta^{\max}_{x_T} M(x(T|u), \widetilde{x}; \delta_{x_T}) + \mathcal{R}^{\max}(u; \beta_{dv}^{\max}, \beta_{dn}^{\max})
	\to \min\limits_{u},   	
	\label{f_22} \\
	\Phi^{\rm Var}(u; \beta^{\rm Var}_{x_T}, \beta^{\rm Var}_{dv}, \beta^{\rm Var}_{dn}) &:=& 
	\beta^{\rm Var}_{x_T} M(x(T|u), \widetilde{x}; \delta_{x_T}) + \mathcal{R}^{\rm Var}(u; \beta^{\rm Var}_{dv}, \beta^{\rm Var}_{dn})
	\to \min\limits_{u}, 
	\label{f_23} \\
	\Phi^{\rm abs}(u; \beta^{\rm abs}_{x_T}, \beta^{\rm abs}_{dv}, \beta^{\rm abs}_{dn}) &:=& 
	\beta^{\rm abs}_{x_T} M(x(T|u), \widetilde{x}; \delta_{x_T}) + \mathcal{R}^{\rm abs}(u; \beta^{\rm abs}_{dv}, \beta^{\rm abs}_{dn})
	\to \min\limits_{u},
	\label{f_24}
\end{eqnarray}  
where the weight coefficients $\beta^{\max}_{x_T}, \beta^{\rm Var}_{x_T}, 
\beta^{\rm abs}_{x_T} >0$, and the parameter $\delta_{x_T} = 0$. Thus, if the constraints (\ref{f_4}) and~(\ref{f_11}) are used, then for any point $\widetilde{x}$ belonging to the RS there should exist such control $u=\widetilde{u}$, which satisfies~(\ref{f_4}),~(\ref{f_11}) and gives zero value for the objective functional in~(\ref{f_22}) including the case 
when the weight coefficients are well balanced. 

\begin{definition}[CSs]
	\label{definition2} 	
	If the system (\ref{f_16}) evolving at a certain range $[0, T]$ is considered with controls which satisfy only the constraint~(\ref{f_4}), then CS $\mathcal{C}(T, x_{\rm target}, \mathcal{U}([0,T],Q))$ for a given target point $x_{\rm target} \in \mathcal{B}$ is the set of all such initial states, $\{x(0)\}$, that for each of them there exists an admissible control $u \in \mathcal{U}([0,T],Q)$ that provides the system's final state $x(T|u)$ coinciding with~$x_{\rm target}$. If the class 
	$\mathcal{U}([0,T],Q)$ consists of controls  satisfying~(\ref{f_4}) and~(\ref{f_11}), then CS is formed by such initial states, $\{x_0\}$, that correspond to the processes $\{x(\cdot),u(\cdot)\}$, each of them satisfies the system of equalities
	\begin{eqnarray}
		M(x(T|u), x_{\rm target}; \delta_{x_T}) = 0, \quad M^{dv}=0, \quad M^{dn} = 0,
		\label{f_25}
	\end{eqnarray}
	where $\delta_{x_T}=0$ is taken. If the regularizer (\ref{f_8}) or (\ref{f_10}) is considered, then CS is formed by the following. For any point $\widetilde{x}$ belonging to the CS there is a control $u=\widetilde{u}$ that  satisfies~(\ref{f_4}) and
	solves, correspondingly, the minimization problem	
	\begin{eqnarray}
		\mathcal{R}^{\rm Var}(u; \beta^{\rm Var}_{dv}, \beta^{\rm Var}_{dn})
		\to \min\limits_{u}, \quad \text{s.t.} \quad M(x(T|u), x_{\rm target}; \delta_{x_T}) = 0
		\label{f_26}	
	\end{eqnarray}
	or the minimization problem 
	\begin{eqnarray}
		\mathcal{R}^{\rm abs}(u; \beta^{\rm abs}_{dv}, \beta^{\rm abs}_{dn})
		\to \min\limits_{u}, \quad \text{s.t.} \quad M(x(T|u), x_{\rm target}; \delta_{x_T}) = 0,    
		\label{f_27}
	\end{eqnarray}
	where $\delta_{x_T} = 0$. For the problems~(\ref{f_26}), (\ref{f_27}), 
control $u$ is considered in the class of controls~(\ref{f_5}), (\ref{f_6}) satisfying only the constraint~(\ref{f_4}). 
\end{definition}

For the problems~(\ref{f_25})--(\ref{f_27}), the following corresponding composite objective functionals to be minimized are suggested:
\begin{eqnarray}
	\Phi^{\max}(u; \beta^{\max}_{x_T}, \beta^{\max}_{dv}, \beta^{\max}_{dn}) 
&:=& 
	\beta^{\max}_{x_T} M(x(T|u), x_{\rm target}; \delta_{x_T}) + \nonumber \\
	&&+ \mathcal{R}^{\max}(u; \beta_{dv}^{\max}, \beta_{dn}^{\max})
	\to \min\limits_{u},  
	\label{f_28} \\
	\Phi^{\rm Var}(u; \beta^{\rm Var}_{x_T}, \beta^{\rm Var}_{dv}, \beta^{\rm Var}_{dn}) &:=& 
	\beta^{\rm Var}_{x_T} M(x(T|u), x_{\rm target}; \delta_{x_T}) + \nonumber \\
	&&+ \mathcal{R}^{\rm Var}(u; \beta^{\rm Var}_{dv}, \beta^{\rm Var}_{dn})
	\to \min\limits_{u},  
	\label{f_29} \\
	\Phi^{\rm abs}(u; \beta^{\rm abs}_{x_T}, \beta^{\rm abs}_{dv}, \beta^{\rm abs}_{dn}) &:=& 
	\beta^{\rm abs}_{x_T} M(x(T|u), x_{\rm target}; \delta_{x_T}) + \nonumber \\
	&&+ \mathcal{R}^{\rm abs}(u; \beta^{\rm abs}_{dv}, \beta^{\rm abs}_{dn})
	\to \min\limits_{u},
	\label{f_30}
\end{eqnarray}
where the weight coefficients $\beta^{\max}_{x_T}, \beta^{\rm Var}_{x_T}, 
\beta^{\rm abs}_{x_T} > 0$, and the parameter $\delta_{x_T} = 0$ is considered.

\section{Definitions and Algorithms for Numerical Estimations of Reachable and Controllability Sets}\label{section5}

For the system~(\ref{f_16}), the problem of estimating its 
RSs and CSs is to  obtain such points in the ball~$\mathcal{B}$, which allow to characterize location, volume of these
RSs and CSs. In~this article, taking into account the articles~\cite{Morzhin_Tyatyushkin_2008, Tyatyushkin_Morzhin_2011}, we define below outer parallelepipedal (interval) estimations and pointwise estimations for RSs and CSs of the system~(\ref{f_16}). The last type of estimations is needed 
for analyzing the interiors of RSs and CSs.

\begin{definition}[outer rectangular estimation for a RS]
	\label{definition3}
	For a RS $\mathcal{R}(T, x_0, \mathcal{U}([0,T],Q))$ of the system (\ref{f_16}), the corresponding outer parallelepipedal estimation $\mathcal{R}_{\square}(T, x_0, \mathcal{U}([0,T],Q))$ is the rectangular parallelepiped defined by the following. If the system (\ref{f_16}) is considered with controls, for which only the constraint~(\ref{f_4}) is used, then 
	$\mathcal{R}_{\square}(T, x_0, \mathcal{U}([0,T],Q))$ is defined by the six coordinates  
	$x_{j,\min}$, $x_{j,\max}$, $j = 1,2,3$, which are obtained by solving 
six variants of the minimization problem
	\begin{eqnarray}
		\Phi_{\square}(u; a) &:=&  \langle a, x(T|u) \rangle \to \min\limits_{u}, 
		\quad a \in \left\{ (\pm 1, 0, 0), (0, \pm 1, 0), (0, 0, \pm 1) \right\}. 
		\label{f_31} 
	\end{eqnarray} 
	If the regularizer (\ref{f_8}) or (\ref{f_10}) or (\ref{f_14}) is considered, then the estimation 
	$\mathcal{R}_{\square}(T, x_0, \mathcal{U}([0,T],Q))$ is defined by the six coordinates  
	$x_{j,\min}$, $x_{j,\max}$, $j = 1,2,3$, which are obtained by solving 
six variants, correspondingly, of the two-criteria minimization problem
	\begin{eqnarray}
		\Phi_{\square}(u; a) =  \langle a, x(T|u) \rangle \to \min\limits_u, \quad 
		\mathcal{R}^{\rm Var}(u; \beta^{\rm Var}_{dv}, \beta^{\rm Var}_{dn})
		\to \min\limits_{u}, 
		\label{f_32}
	\end{eqnarray} 
	or the two-criteria minimization problem
	\begin{eqnarray}
		\Phi_{\square}(u; a) =  \langle a, x(T|u) \rangle \to \min\limits_u, \quad  
		\mathcal{R}^{\rm abs}(u; \beta^{\rm abs}_{dv}, \beta^{\rm abs}_{dn})
		\to \min\limits_{u},  
		\label{f_33}
	\end{eqnarray}
	or the minimization problem
	\begin{eqnarray}
		\Phi_{\square}(u; a) = 
		\langle a, x(T|u) \rangle \to \min\limits_u, \quad \text{s.t.} \quad M^{dv} = 0, \quad M^{dn} = 0.
		\label{f_34}
	\end{eqnarray} 
\end{definition}

For the minimization problems~(\ref{f_32})--(\ref{f_34}), the following corresponding composite objective functionals to be minimized are formulated:
\begin{eqnarray}
	\Phi_{\square}^{\rm Var}(u; a, \beta^{\rm Var}_{\square, x_T}, \beta^{\rm Var}_{dv}, \beta^{\rm Var}_{dn}) &:=& 
	\beta^{\rm Var}_{\square, x_T} \Phi_{\square}(u; a) + \mathcal{R}^{\rm Var}(u; \beta^{\rm Var}_{dv}, \beta^{\rm Var}_{dn})
	\to \min\limits_{u}, 
	\label{f_35} \\
	\Phi_{\square}(u; a, \beta^{\rm abs}_{\square, x_T}, \beta^{\rm abs}_{dv}, \beta^{\rm abs}_{dn}) &:=& 
	\beta^{\rm abs}_{\square, x_T} \Phi_{\square}(u; a) + \mathcal{R}^{\rm abs}(u; \beta^{\rm abs}_{dv}, \beta^{\rm abs}_{dn})
	\to \min\limits_{u}, 
	\label{f_36} \\
	\Phi_{\square}^{\max}(u; a, \beta^{\max}_{\square, x_T}, \beta^{\max}_{dv}, \beta^{\max}_{dn}) &:=& 
	\beta_{\square, x_T}^{\max} \Phi_{\square}(u; a) + \mathcal{R}^{\max}(u; 
\beta_{dv}^{\max}, \beta_{dn}^{\max})
	\to \min\limits_{u},
	\label{f_37}
\end{eqnarray} 
where the weight coefficients
$\beta^{\rm Var}_{\square, x_T}, \beta^{\rm abs}_{\square, x_T}, \beta^{\max}_{\square, x_T} > 0$.

\begin{figure}[h!]
	\centering
	\includegraphics[width=1\linewidth]{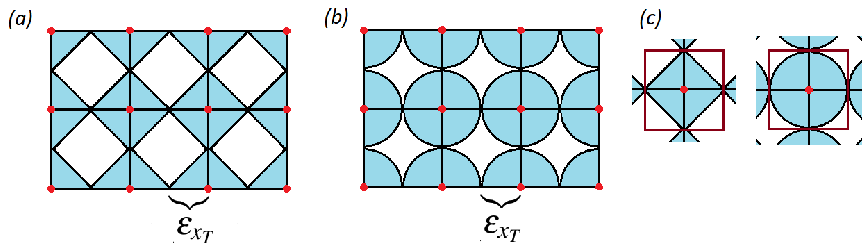} 
	\caption{$(\varepsilon_{x_T}/z)$-networks and a particular cubes, where $z=1$, in their intersections with a coordinate plane: (a)~network for $p=1$; (b)~network for $p=2$; (c)~cubes with an edge length equal to $2\varepsilon_{x_T}$. 
		A~node $x^s$ (see Definition~\ref{definition4}) is shown via circle marker. The filled areas indicate the points $\{x\}$ satisfying the condition~(\ref{f_39}) with~$z=1$.}
	\label{Fig1}
\end{figure}

In the Bloch ball consider the uniform grid
\begin{eqnarray} 
	G(M) := \Big\{ x^s = (x_1^s, x_2^s, x_3^s) \in [-1, 1]^3 : 
	\quad (x_j^s)^i = -1 + \frac{2}{M} i, \nonumber \\
	j = \overline{1,3}, \quad i = \overline{0,M}, \quad 
	s = \overline{1, (M+1)^3} \Big\} \bigcap \mathcal{B},
	\label{f_38}
\end{eqnarray}
where the discretization step $2/M$ is defined by some natural number $M$. If the step $2/M$ in (\ref{f_32}) is, e.g., $1/10$, then the grid $G(M)$ is formed by 4169 nodes. 
Consider the inequality 
\begin{eqnarray}  \left\| x - \widehat{x} \right\|_p^p \leq \left( \delta_{x_T} \right)^p, \qquad p \in \{1, 2\}, 
	\label{f_39}
\end{eqnarray}
where we set $\delta_{x_T} = \varepsilon_{x_T}/z > 0$, $\varepsilon_{x_T} = 1/M$, i.e. $\delta_{x_T} = 1/(Mz)$, and the parameter $z \in [1, z_{\max}]$ is introduced for additional regulating 
the accuracy of reachability. The grid (\ref{f_38}) and the inequality (\ref{f_39}) define the  $(\varepsilon_{x_T}/z)$-networks. Fig.~\ref{Fig1}(a,b) schematically illustrates these networks, which correspond 
to $p = 1,2$ and $z=1$, in their intersections with a coordinate plane.

\begin{definition}[pointwise estimation for a RS]
	\label{definition4}
	For a RS $\mathcal{R}(T, x_0, \mathcal{U}([0,T],Q))$ of the system (\ref{f_16}), the corresponding pointwise estimation $\widehat{\mathcal{R}}(T, 
x_0, \mathcal{U}([0,T],Q))$ is defined by the following. If only the constraint~(\ref{f_4}) is considered for defining the class of controls, then 
$\widehat{\mathcal{R}}(T, x_0, \mathcal{U}([0,T],Q))$ is formed by all such the endpoints $\{x(T|u)\} \subset \mathcal{R}(T, x_0, \mathcal{U}([0,T],Q))$  
	that each of these points satisfies the condition (\ref{f_39}) with $x = 
x(T|u)$, $\widehat{x} = x^s$. In other words, here each point $\widetilde{x} \in \widehat{\mathcal{R}}(T, x_0, \mathcal{U}([0,T],Q))$ is an endpoint of the trajectory $x(\cdot) = \widetilde{x}(\cdot)$  representing the solution of the minimization problem
	\begin{eqnarray} 
		\Phi(u|~x^s, \delta_{x_T}) := M \left( x(T|u), x^s; \delta_{x_T} \right) \to \min\limits_{u}  
		\label{f_40}
	\end{eqnarray}
	for the node $x^s \in G(M)$, which is nearest to the point 
	$\widetilde{x}$, where $\delta_{x_T} = 1/(Mz)$. If the regularizer (\ref{f_8}) or (\ref{f_10}) or (\ref{f_14}) is considered, then the estimation 
	$\widehat{\mathcal{R}}(T, x_0, \mathcal{U}([0,T],Q))$ is defined by the endpoints $\{x(T|u)\} \subset \mathcal{R}(T, x_0, \mathcal{U}([0,T],Q))$, 
each of them is obtained by solving, correspondingly, (\ref{f_19}) or (\ref{f_20}) or (\ref{f_21}) with $\delta_{x_T} = 1/(Mz)$.
\end{definition} 

For the problems~(\ref{f_19}), (\ref{f_20}), and (\ref{f_21}), 
the corresponding minimization problems (\ref{f_28}), (\ref{f_29}), and (\ref{f_30}) with $\delta_{x_T} = 1/(Mz)$ are considered. 
In contrast to Definition~\ref{definition1}, we consider approximate reachability in Definition~\ref{definition4} in the terms of the threshold~$\delta_{x_T} > 0$. For a 
RS, its volume is approximately equal 
to the sum of all particular cubes (see Fig.~\ref{Fig1}(c)), which centers are such that their vicinities defined by~(\ref{f_39}) contain the system's endpoints $\{x(T)\}$. The volume of each particular cube is equal to~$(2\varepsilon_{x_T})^{3} = 8/M^3$ (for example, if $M=20$, then $\varepsilon_{x_T}=0.05$ and the volume of a particular cube is equal to 0.001). 

The results of solving the  problems~(\ref{f_22})--(\ref{f_24}), (\ref{f_28})--(\ref{f_30}), (\ref{f_35})--(\ref{f_37}) depend on the weight parameters of the objective functionals used in these problems. For example, consider the regularizer~(\ref{f_8}) and the minimization problems (\ref{f_35}) and (\ref{f_29}) with $\delta_{x_T} = 1/(Mz)$. Both for (\ref{f_35}) and (\ref{f_29}), consider the same weight parameters of the regularizer (\ref{f_8}). Consider some value $\beta^{\rm abs}_{\square, x_T} = \beta^{\rm abs}_{x_T}$ both in (\ref{f_35}) and (\ref{f_29}). In this case, in general, the meanings of the weight coefficients are different, because $\langle a, x(T) \rangle$ can be equal, e.g., to~1, while $M(x(T|u), x^s, \delta_{x_T})$ can achieve zero. That is why, for obtaining the pointwise estimation for a RS with using (\ref{f_29}) it can be better 
to base on the RS's outer parallelepipedal estimation found with taking into account only the constraint~(\ref{f_4}). However, obtaining outer parallelepipedal estimations using (\ref{f_32})--(\ref{f_37}) has an independent interest.

For CSs of the system~(\ref{f_16}), their outer parallelepipedal and pointwise estimations are defined by analogy with Definitions~\ref{definition3},~\ref{definition4}. Below we write the definition, e.g., of outer parallelepipedal estimation of a CS, when only the constraint~(\ref{f_4}) is used. 

\begin{definition}[outer rectangular estimation for a CS, no additional constraints on controls]
	\label{definition5}
	For a CS $\mathcal{C}(T, x_{\rm target}, \mathcal{U}([0,T],Q))$ of the system (\ref{f_16}) considered with controls satisfying only the constraint~(\ref{f_4}), 
	the corresponding outer parallelepipedal estimation 
	$\mathcal{C}_{\square}(T, x_{\rm target}, \mathcal{U}([0,T],Q))$ is the rectangular parallelepiped defined  
	by the six coordinates $x_{0,j,\min}$, $x_{0,j,\max}$, $j = 1,2,3$, which are obtained by solving the following minimization problem for each $a \in \left\{ (\pm 1, 0, 0), (0, \pm 1, 0), (0, 0, \pm 1) \right\}$:
	\begin{eqnarray}
		\Phi_{\square}(u, p; a) &:=&  \langle a, p \rangle \to \min\limits_{(u,p)} \qquad \text{s.t.} \quad x(0)=p, \quad x(T|u)=x_{\rm target},  
		\label{f_41} 
	\end{eqnarray} 
	where the controlling vector parameter $p=(p_1, p_2, p_3)$ s.t. $p_1^2 
+ p_2^2 + p_3^2 \leq 1$. 
\end{definition}

For the problem~(\ref{f_41}), one can consider the following composite functional to be minimized:
\begin{eqnarray}
	\Phi_{\square}(u, p; a, \beta_{x_0}, \beta_{x_T}) := \beta_{x_0} \langle a, p \rangle + \beta_{x_T} M\left(x(T|u), x_{\rm target}, \delta_{x_T} 
\right),
	\label{f_42}
\end{eqnarray}
where the weight coefficients $\beta_{x_0}, \beta_{x_T} > 0$, and the value $\delta_{x_T} = 1/(Mz)$. 

\begin{definition}[pointwise estimation of a CS]
	\label{definition6}
	For a CS $\mathcal{C}(T, x_{\rm target}, \mathcal{U}([0,T],Q))$ of the system (\ref{f_16}), the corresponding pointwise estimation $\widehat{\mathcal{C}}(T, x_{\rm target}, \mathcal{U}([0,T],Q))$ is defined by the following. If~only the constraint~(\ref{f_4}) is considered, then $\widehat{\mathcal{C}}(T, x_{\rm target}, \mathcal{U}([0,T],Q))$ is formed by all such
	nodes $\{ x^s\}$ of the grid $G(M)$ that each of these nodes is an initial point of the trajectory representing the solution of the minimization problem 
	\begin{eqnarray} 
		\Phi(u|~x_{\rm target}, \delta_{x_T}) := M \left( x(T|u), x_{\rm target}; \delta_{x_T} \right) \to \min\limits_{u} 
		\label{f_43}
	\end{eqnarray} 
	with $\delta_{x_T}=1/(Mz)$. If the regularizer (\ref{f_8}) or (\ref{f_10}) or (\ref{f_14}) is considered, then the estimation
	$\widehat{\mathcal{C}}(T, x_{\rm target}, \mathcal{U}([0,T],Q))$ is defined by such nodes $\{x^s\}$ of the grid $G(M)$ that each of them is an initial point of the trajectory representing the solution, correspondingly, 
of (\ref{f_25}) or (\ref{f_26}) or (\ref{f_27}) with $\delta_{x_T} = 1/(Mz)$. 
\end{definition}   

With respect to the interest how changing the bounds $v_{\min}$, 
$v_{\max}$, $n_{\max}$ in (\ref{f_4}) modifies estimations of RSs and CSs, consider the classes   
\begin{eqnarray} 
	\mathcal{U}([0,T],Q^q), 
	\quad Q^q:= \left[v_{\min} d^q, v_{\max} d^q\right] 
	\times \left[0, n_{\max} d^q \right]
	\label{f_44} 
\end{eqnarray} 
generated by such a multiplier 
$d^q$ that $(d^q)_{q=1}^7 = \left( 1,~ 0.8,~ 0.6,~ 0.4,~ 0.2,~ 0.1,~ 0.05 \right)$. The case $q=1$ gives $Q$ defined in (\ref{f_4}). Here we 
consider the vector 
\begin{eqnarray}
	{\bf u} &=& \left( \left\{ v_j \right\}_{j=0}^{N_v-1}, 
	\left\{ n_j \right\}_{j=0}^{N_n-1} \right) \in Q^q(N_v, N_n) := \left[v_{\min} d^q, v_{\max} d^q\right]^{N_v} \times 
	\left[0, n_{\max} d^q\right]^{N_n} \subseteq \nonumber \\ 
	&&\subseteq  Q(N_v,N_n)  \subset
	\mathbb{R}^{N_v + N_n},  
	\label{f_45}
\end{eqnarray}
where $(N_v + N_n)$-dimensional search spaces $Q^q(N_v, N_n)$ are defined 
for different $q$.   

\begin{algorithm}[estimating RSs with/ without considering any regularizer~(\ref{f_8})/ (\ref{f_9})/~(\ref{f_14})]
	\label{algorithm1} 
	{\rm 
	Estimating the sets $\mathcal{R}(T, x_0, 
	\mathcal{U}([0,T], Q^q))$, $q = \overline{1,7}$, of the system (\ref{f_16}) using the classes~(\ref{f_44}) with/ without any regularizer~(\ref{f_8})/(\ref{f_9})/~(\ref{f_14}). Set $p \in \{1,2\}$. At~the $q$th iteration, the following operations are evaluated.
	
	\par \underline{Step 1}. Find  the outer parallelepipedal estimation
	$\mathcal{R}_{\square}(T, x_0, \mathcal{U}([0,T], Q^q))$ by globally solving six one-type OCPs~(\ref{f_31}), where control $u$ is considered in the class~$\mathcal{U}([0,T], Q^q))$, $q \in \{1,2,..., 7\}$, defined using only the constraint~(\ref{f_4}).  
	
	\par \underline{Step 2}. If $q=1$, then find the set $G_{\square}^q(M)$ formed by all such nodes of $G(M)$ that are bounded by the parallelepiped $\mathcal{R}_{\square}(T, x_0, \mathcal{U}([0,T],Q^1))$. If~$q>1$, then find the set $G_{\square}^q(M)$, which is formed by all such nodes 
	of the set $G^q_{\widehat{\mathcal{R}}}(M)$
	(this set is defined below, in the 3rd step, and is known here when $q>1$) that are bounded by $\mathcal{R}_{\square}(T, x_0, \mathcal{U}([0,T],Q^q))$.
	
	\par \underline{Step 3}. This step is for checking, whether a node $x^s \in G_{\square}^q(M)$, where $s \in \overline{1, {\rm card}(G_{\square}^q(M))}$ 
	(here ``card'' mean ``cardinality''), is reachable 
	from the given initial state $x_0$. If the class $\mathcal{U}([0,T];Q^q)$ is defined with only the constraint~(\ref{f_4}), then the pointwise estimation of the RS is obtained by solving the series of such OCPs, each of 
them is of the type~(\ref{f_40}) and is for checking reachability of a node $x^s \in G_{\square}^q(M)$  in the meaning (\ref{f_39}) (see Fig.~\ref{Fig1}), where $x = x(T|u)$ and $\widehat{x} = x^s$ are taken. 
	
	If the class $\mathcal{U}([0,T];Q^q)$ is defined with the constraints~(\ref{f_4}) and~(\ref{f_11}), then the pointwise estimation of the RS is computed by solving the series of such problems, each of them is of the type (\ref{f_19}) with $\delta_{x_T} = 1/(Mz)$. For each problem of the type~(\ref{f_19}), the corresponding minimization problem~(\ref{f_22}) is used. For the whole series of the problems of the type~(\ref{f_22}), 
we set some values for the weight coefficients $\beta_{x_T}^{\max}, \beta_{dv}^{\max}, \beta_{dn}^{\max} > 0$ by looking for some balance between the three terms in the objective functional. If the constraint~(\ref{f_4}) and the regularizer~(\ref{f_8})/ (\ref{f_10}) are used, then the pointwise estimation of the RS is computed by solving the series of such OCPs, each of them is of the type (\ref{f_20})/ (\ref{f_21}) with $\delta_{x_T} 
= 1/(Mz)$. For each OCP of the mentioned type
	(\ref{f_20})/ (\ref{f_21}), the corresponding OCP of the type (\ref{f_23})/ (\ref{f_24}) is considered. For the whole series of OCPs of the type (\ref{f_23})/ (\ref{f_24}), we set some values of the corresponding weight coefficients. 
	
	For any of the mentioned four cases, we divide the corresponding series of the OCPs into some number of batches for parallel computations. For a node $x^s$, if several runs of DEM and/or DAM do not allow to classify this node as approximately reachable, then the node is mentioned as unreachable. As the result, we form the set $\widehat{\mathcal{R}}(T, x_0, \mathcal{U}([0,T],Q^q))$ 
	of all selected endpoints, $\{ x(T)\}$, and the set $G^{q}_{\widehat{\mathcal{R}}}(M)$ of the corresponding nodes of the grid~$G(M)$.    
}
\end{algorithm}

In the terms of the algorithm's complexity it is important to use outer parallepipedal estimations and taking into account the fact that the RS
$\mathcal{R}(T, x_0, \mathcal{U}([0,T], Q^{q-1})$ includes the RS
$\mathcal{R}(T, x_0, \mathcal{U}([0,T], Q^q))$, $q \geq 2$. 

\begin{algorithm}[estimating CSs with/ without considering any regularizer~(\ref{f_8})/ (\ref{f_9})/~(\ref{f_14})]
	\label{algorithm2} 
	{\rm 
	Estimating the sets $\mathcal{C}(T, x_{\rm target}, 
	\mathcal{U}([0,T], Q^q))$, $q = \overline{1,7}$, of the system (\ref{f_16}) using the classes~(\ref{f_44}) with/ without any  regularizer~(\ref{f_8})/(\ref{f_9})/~(\ref{f_14}). Set $p \in \{1,2\}$. At~the $q$th iteration, the following operations are carried out. 
	
	\par \underline{Step 1}. Find  the outer parallelepipedal estimation
	$\mathcal{C}_{\square}(T, x_{\rm target}, \mathcal{U}([0,T], Q^q))$ by globally solving six one-type OCPs~(\ref{f_41}) (also see (\ref{f_42})), where control $u$ is considered in the class~$\mathcal{U}([0,T], Q^q))$, $q \in \{1,2,..., 7\}$, defined using only the constraint~(\ref{f_4}). 
	
	\par \underline{Step 2}. If $q=1$, then find the set $G_{\square}^q(M)$ formed by all such nodes of $G(M)$ that are bounded by the parallelepiped $\mathcal{C}_{\square}(T, x_{\rm target}, \mathcal{U}([0,T],Q^1))$. If~$q>1$, then find the set $G_{\square}^q(M)$, which is formed by all such 
nodes 
	of the set $G^q_{\widehat{\mathcal{C}}}(M)$
	(this set is defined below, in the 3rd step, and is known here when $q>1$) that are bounded by $\mathcal{C}_{\square}(T, x_{\rm target}, \mathcal{U}([0,T],Q^q))$.

	\par \underline{Step 3}. This step is for checking, whether a node $x^s \in G_{\square}^q(M)$, where $s \in \overline{1, {\rm card}(G_{\square}^q(M))}$, can be as initial state $x_0$ in (\ref{f_16}) for moving the system to the given target state~$x_{\rm target}$. If the class $\mathcal{U}([0,T];Q^q)$ is defined with only the constraint~(\ref{f_4}), then the pointwise estimation of the RS is obtained by solving the series of such OCPs, each of them is of the type~(\ref{f_43}) with $\delta_{x_T}=1/(Mz)$ and is for checking a node $x^s \in G_{\square}^q(M)$ to be such an 
initial state that the system can be  moved (approximately) to the given $x_{\rm target}$ in the meaning of the inequality (\ref{f_39}), where $x = x(T|u)$ and $\widehat{x} = x_{\rm target}$ are taken. 
	
	If the class $\mathcal{U}([0,T];Q^q)$ is defined with the constraints~(\ref{f_4}) and~(\ref{f_11}), then the pointwise estimation of the CS is computed by solving the series of such problems, each of them is of the type (\ref{f_25}) with $\delta_{x_T} = 1/(Mz)$. For each problem of the type~(\ref{f_25}), the corresponding minimization problem~(\ref{f_28}) is used. For the whole series of the problems of the type~(\ref{f_28}), 
some values for the weight coefficients $\beta_{x_T}^{\max}, \beta_{dv}^{\max}, \beta_{dn}^{\max} > 0$ are set by looking for  some balance between the three terms in the objective functional. If the constraint~(\ref{f_4}) and the regularizer~(\ref{f_8})/ (\ref{f_10}) are used, then the pointwise estimation of the RS is computed by solving the series of such OCPs, each of them is of the type (\ref{f_26})/ (\ref{f_27}) with $\delta_{x_T} = 1/(Mz)$. For each OCP of the mentioned type (\ref{f_26})/ (\ref{f_27}), the corresponding OCP of the type (\ref{f_29})/ (\ref{f_30}) is considered. For the whole series of OCPs of the type (\ref{f_29})/ (\ref{f_30}), some values of the corresponding weight coefficients are set. 
	
	For any of the mentioned four cases, the corresponding series of the OCPs is divided into some number of batches for parallel computations. For a 
node $x^s$, if several runs of DEM and/or DAM do not allow to classify this node as belonging to the CS, then this node is mentioned as beyond the 
CS. As the result, we form the set 
	$G^{q}_{\widehat{\mathcal{C}}}(M)$ of all selected nodes. This set is taken as  
	$\widehat{\mathcal{C}}(T, x_{\rm target}, \mathcal{U}([0,T],Q^q))$.
}
\end{algorithm}

\begin{statement}[upper bounds]
	\label{statement1} 	
	Consider controls (\ref{f_5}), (\ref{f_6}) satisfying the constraint~$u(t) = (v(t),n(t)) \in Q^q$. There are the following upper bounds:
	\par --- for the variations (\ref{f_9}):
	\begin{eqnarray}
		{\rm Var}_{[0,T]}(v; N_v) \leq  d^q\left(v_{\max} - v_{\min} \right)\left( N_v - 1 \right), \quad 
		{\rm Var}_{[0,T]}(n; N_n) \leq d^q n_{\max} \left( N_n - 1 \right);
		\label{f_46}
	\end{eqnarray} 
	\par --- for the sums used in~(\ref{f_10}):
	\begin{eqnarray}
		\sum\limits_{j=0}^{N_v-1} |v_j| \leq d^q \max\{|v_{\min}|, v_{\max}\} 
N_v, \qquad  \sum\limits_{j=0}^{N_n-1} n_j \leq d^q n_{\max} N_n;
		\label{f_47}
	\end{eqnarray} 
	\par --- for $M^{dv}$, $M^{dn}$, $M^{\delta_{dv}}$, and $M^{\delta_{dv}}$ defined in (\ref{f_12}) and~(\ref{f_13}):
	\begin{eqnarray}
		M^{dv} &\leq& d^q \left( v_{\max} - v_{\min} \right), \qquad
		M^{\delta_{dv}} \leq \max\left\{ d^q \left( v_{\max} - v_{\min} \right) 
- \delta_{dv}, 0 \right\}, 
		\label{f_48} \\
		M^{dn} &\leq& d^q n_{\max}, \qquad \qquad \quad~~ 
		M^{\delta_{dn}} \leq \max\left\{ d^q n_{\max} - \delta_{dn}, 0 \right\}.
		\label{f_49}	
	\end{eqnarray} 
\end{statement} 

The upper bounds given in (\ref{f_46})--(\ref{f_49}) can be used for adjusting the weight coefficients in the minimization problems used in Algorithms~\ref{algorithm1},~\ref{algorithm2}.  


\section{Using Stochastic Zeroth-Order Optimization Methods}\label{section6}

Because the considered above OCPs are also the finite-dimensional minimization problems due to piecewise constant type of controls $v,n$, we use DEM and DAM directly to these OCPs, in contrast to the approach of reduction an OCP to finite-dimensional optimization by approximating piecewise continuous controls by piecewise constant controls~\cite{Morzhin_Pechen_LJM_2019, Morzhin_Pechen_Physics_of_Particles_and_Nuclei}. 

DEM and DAM are based on some heuristic strategies for searching approximations 
for the global minimum of an objective function. These methods can be applied for implicitly defined, multi-modal, non-differentiable objective functions. Taking $p=1$ in the given above objective functionals, which use~(\ref{f_18}), we have the problems for minimizing the non-differentiable objective functions. Taking into account the stochastic nature (automatically generated values of the stochastic variables) of DEM and DAM, it is suggested to make several runs of DEM and/or DAM for the same minimization problem.
Moreover, it is possible to change such non-stochastic variables in DAM as the initial ``temperature''. Of course, if we solve the problem (\ref{f_22})/  (\ref{f_28})/ (\ref{f_40})/ (\ref{f_43}) and obtain zero value of 
the corresponding objective function in the first  run of DEM or DAM, then the problem has been solved and we stop the computations. An example of 
another situation gives the problem (\ref{f_35}), where the composite objective function consists of the three terms: $\beta^{\rm Var}_{\square, x_T} \langle a, x(T) \rangle$ and two terms of the regularizer~(\ref{f_8}). In such situation, it is logical to make several runs of DEM and/ or DAM for further comparing different results. Although the problem (\ref{f_22})/ (\ref{f_28})   considers $M(x(T|u), \widetilde{x}; \delta_{x_T})$, 
$M(x(T|u), x_{\rm target}; \delta_{x_T})$, $M^{dv}, M^{dn}$, each of them 
has to reach zero, it is also important to set the weight coefficients by 
looking for some balance between the terms, which are in the composite functional in (\ref{f_22})/ (\ref{f_28}), for avoiding early stop of some optimization algorithm process in the situation, when the priority of some 
term is low than the sensitivity threshold used for stopping in the algorithm. 
The complexity of the approach using DEM and DAM depends mainly on the dimension $(N_v+N_n)$ of the search space $Q^q(N_v,N_n)$. The numbers $N_v$, $N_n$ have to be taken to satisfy some trade-off between having a small 
time step $\Delta t  = T/N_v = T/N_n$ and working with DEM and/ or DAM in a reasonably low dimensional search space~$Q^q(N_v,N_n)$.   


\section{Numerical Results}\label{section7}

This section describes our numerical results for estimating RSs and CSs of the system~(\ref{f_16}). These results were obtained using the  Python~3 programs written by the first author. These programs use: (a)~the implementation~\cite{scipy_differential_evolution} of DEM and implementation~\cite{scipy_dual_annealing} of DAM available in {\tt SciPy} scientific computing library; (b)~the tool {\tt odeint}~\cite{scipy_odeint} available in SciPy (as it is noted in~\cite{scipy_odeint}, {\tt odeint} represents {\tt lsoda} from the FORTRAN library {\tt odepack}); (c)~the tool {\tt sqlite3}~\cite{sqlite3_python} for storing the numerical results in SQLite database format; (d)~some another well-known tools for Python~3 programming. Parallel computations were organized as Algorithms~\ref{algorithm1},~\ref{algorithm2} suggest. {\tt odeint} was used for accurate integration of the dynamical system with some given piecewise constant controls~$v,n$.

\subsection{ Without Additional Constraints on Controls}\label{subsection7.1} 

The system (\ref{f_16}) is considered here for the following arbitrary taken values of its parameters: $\omega = 1$, $\kappa = 0.01$, $\gamma = 0.05$. We set $v_{\min} = -100$, $v_{\max} = 100$, and $n_{\max} = 20$ in~(\ref{f_4}), i.e. for $q = 1$ in (\ref{f_44}). For analyzing, how the system's RSs can depend on selecting initial state and final time, we considered $x_0 \in \{ (0.5, 0, 0),~ (0, 0, 1),~ (0, 0, 0) \}$ and 
$T \in \{ 5, 10, 20 \}$. These cases for $x_0$ are significantly different: the point $(0,0,0)$ represents the center of the Bloch ball and the completely mixed quantum state; the point $(0,0,1)$ represents a pole of the Bloch ball (some pure quantum state); the point $(0.5, 0, 0)$ is inside 
the Bloch ball equidistantly from the center of the ball and from the Bloch sphere. Here, for each $T$, we consider $3 \times 7 = 21$~RSs. Thus, 
we considered the problem of estimating $21 \times 3 = 63$~RSs. 

\begin{figure}[ht!]
	\centering
	\includegraphics[width=1\linewidth]{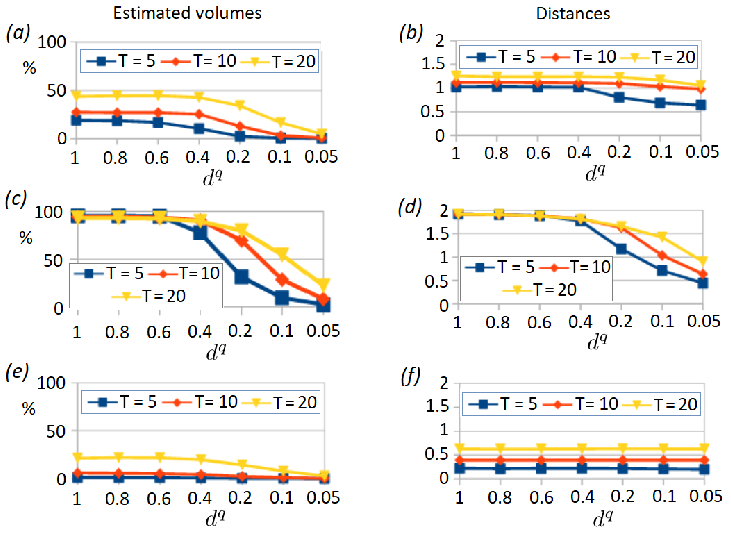} 
	\caption{For the RSs related to 
		$x_0 \in \{ (0.5,0,0), (0,0,1), (0,0,0)\}$, $T \in \{ 5, 10, 20 \}$, 
		and $d^q \in \{1, 0.8, 0.6, 0.4, 0.2, 0.1, 0.05 \}$, 
		the characteristics of the pointwise estimations obtained 
		using Algorithm~\ref{algorithm1}. Estimated 
		volumes of the RSs: (a)~for $x_0 = (0.5, 0, 0)$; (c)~for 
		$x_0 = (0, 0, 1)$; (e)~for $x_0 = (0, 0, 0)$. For each 
		$x_0$, the distance between $x_0$ and the farthest endpoint 
		in the corresponding pointwise estimation: (b)~for 
		$x_0 = (0.5, 0, 0)$; (d)~for $x_0 = (0, 0, 1)$; 
		(f)~for $x_0 = (0, 0, 0)$.}
	\label{Fig2}
\end{figure}

\begin{figure}[ht!]
	\centering
	\includegraphics[width=1\linewidth]{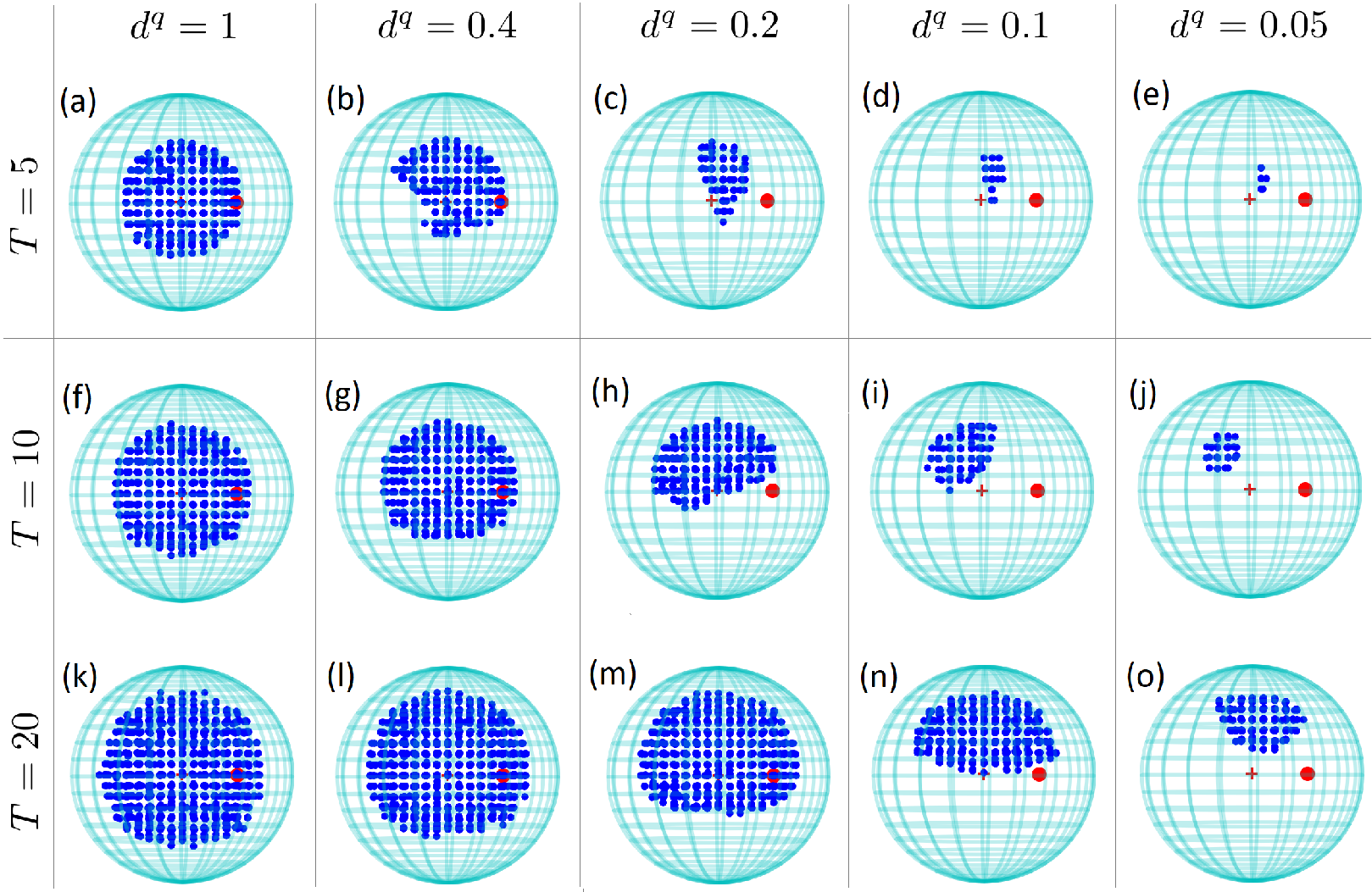}
	\caption{Estimations obtained using Algorithm~\ref{algorithm1}
		for $x_0 = (0.5, 0, 0)$, $T \in \{5, 10, 20\}$, 
		$d^q \in \{1, 0.4, 0.2, 0.1, 0.05 \}$. 
		Round red marker indicates the point $x_0$.}
	\label{Fig3}
\end{figure}

\begin{figure}[ht!]
	\centering
	\includegraphics[width=1\linewidth]{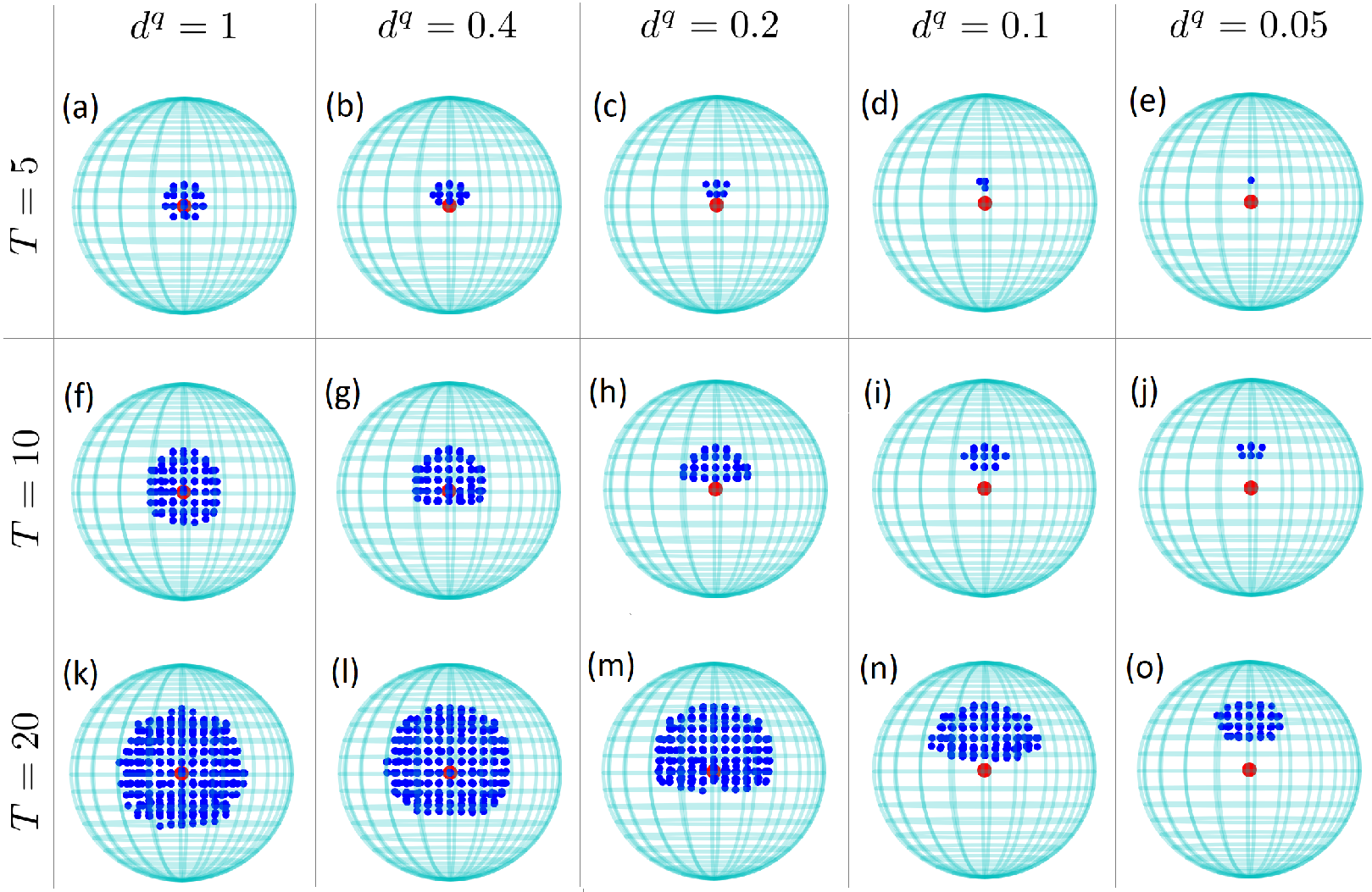} 
	\caption{Estimations computed using Algorithm~\ref{algorithm1} 
		for $x_0 = (0, 0, 0)$, $T \in \{5, 10, 20\}$, 
		and $d^q \in \{1, 0.4, 0.2, 0.1, 0.05 \}$. 
		Round red marker indicates the point $x_0$.}
	\label{Fig4}
\end{figure}

\begin{figure}[ht!]
	\centering
	\includegraphics[width=1\linewidth]{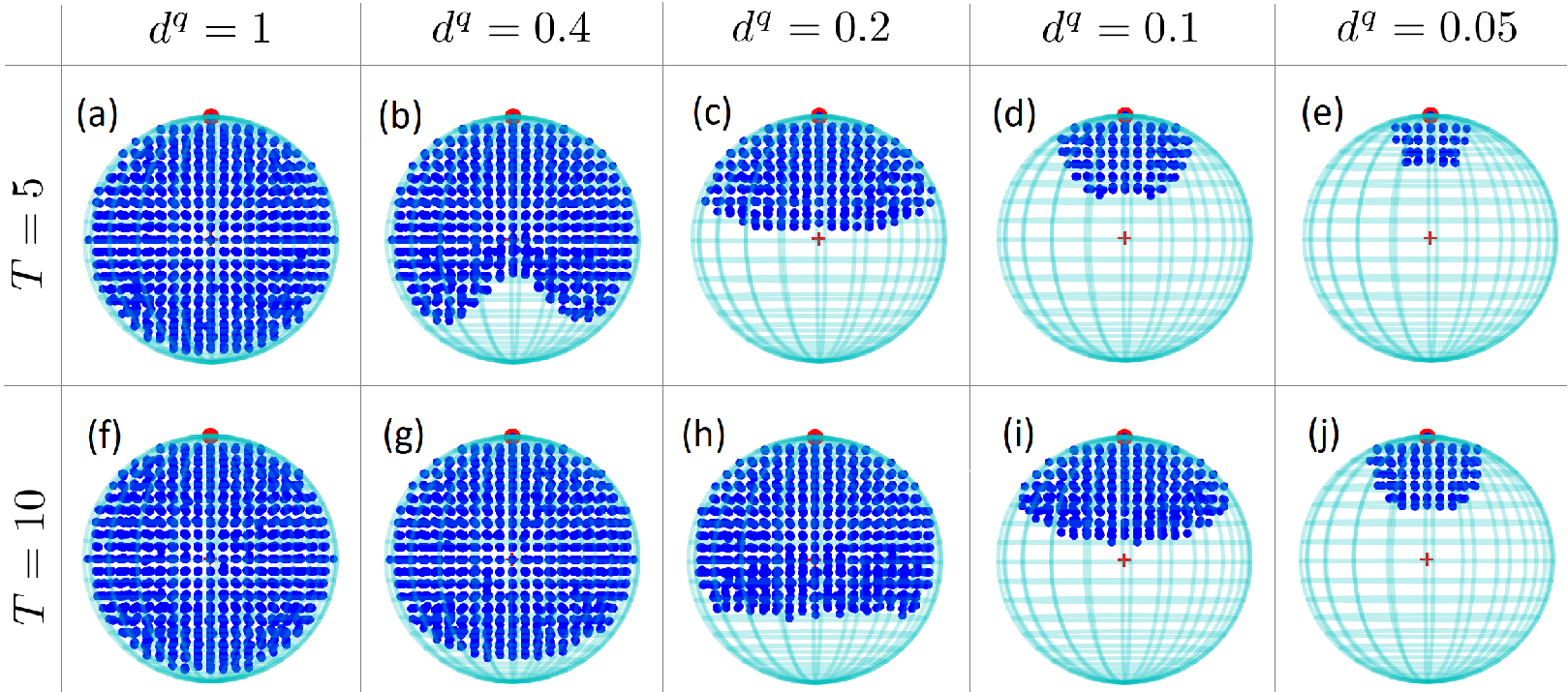} 
	\caption{Estimations found using Algorithm~\ref{algorithm1}  
		for $x_0 = (0, 0, 1)$, $T \in \{5, 10\}$, and 
		$d^q \in \{1, 0.4, 0.2, 0.1, 0.05 \}$. 
		Round red marker indicates the point $x_0$.}
	\label{Fig5}
\end{figure}

Using Algorithm~\ref{algorithm1} with DAM, we numerically estimated
the mentioned 63 RSs. Here  $\varepsilon_{x_T} = 0.05$ was set in~(\ref{f_38}). For each RS $\mathcal{R}(T,x_0, \mathcal{U}([0,T],Q^q))$, its volume 
is estimated by the formula $(2\varepsilon_{x_T})^{3} 
{\rm card}(\widehat{\mathcal{R}}(T, x_0, \mathcal{U}\left([0, T]; Q^q\right)))$
(see Fig.~\ref{Fig1}(c)). The volume of the Bloch ball is equal to $4\pi/3$. For $\varepsilon_{x_T} = 0.05$, the volume of each particular cube is $(2\varepsilon_{x_T})^3 = 0.001$. 
For each 
$x_0 \in \{ (0.5, 0, 0), (0, 0, 1), (0,0,0) \}$, Fig.~\ref{Fig2} 
shows (1)~estimated volumes of the RSs and (2)~distances between 
the initial state  and the maximally distant points of the 
corresponding RSs. In~Fig.~\ref{Fig2}, we see that, for the same 
$T$, decreasing $d^q$ can significantly decrease the estimated volumes of 
the RSs and the distances between the initial states and the maximally distant points 
of the corresponding pointwise estimations. The estimations related to $x_0 = (0,0,1)$ 
are essentially different than the estimations corresponding to $x_0 = (0,0,0)$. 
For instance, if $x_0 = (0,0,1)$ and $d^q =1$, the estimated volume 
of the RS even for $T=5$ is almost equal to the Bloch ball's volume; 
however, for $x_0 = (0,0,0)$, the volume is near 1.1 \% for the same $d^q$ and~$T$. 

Among these 63 estimations, 
Fig.~\ref{Fig3}--\ref{Fig5} show 40~estimations. All 40~estimations are vizualized with zero inclination angle 
of the $x_3$-axis and with the same rotation angle. These visializations were made by the tool {\tt Matplotlib}~\cite{Matplotlib} 
applied to our numerical results. We see that decreasing $d^q$ leads to obtaining 
the pointwise estimations of rather different forms shown in  
Fig.~\ref{Fig3}--\ref{Fig5}, e.g., like to ball or 
a half of ball. As Fig.~\ref{Fig5}(b) clearly shows, for $T=5$ and $d^q 
= 0.4$, 
the corresponding RS is estimated as not convex. Some estimation 
are situated in all 8~orthants. The figures illustrate that decreasing $T$ and $d^q$ can give essentially decrease the estimated volumes; in other 
words, control possibilities are changed.

As Fig.~\ref{Fig3}--\ref{Fig5} show, for some fixed $x_0$ and $d^q$,  it is possible that 
the system's RSs, which relate to different values of $T$, contain the same point. 
This fact relates to the problem of moving the system from $x_0$ to $x_{\rm target}$ 
with minimizing the final time. For each $x_0$, estimating the system's RSs for 
the sequential final times $T = 5, 10, 20$ gives, in other words, three 
time 
sections of the reachable 
tube $\mathcal{R}((0,20]; x_0, \mathcal{U}([0,T];Q^q)) \subset (0,T] \times \mathcal{B}$ as some its estimation. 


\newpage

\subsection{\label{subsection7.2} With Additional Constraints on Controls}

{\bf Estimating RSs of the system (\ref{f_16}) in the situation when the regularizer (\ref{f_8}) is used.} For an illustration, consider the system (\ref{f_16}) with $\kappa = 0.01$, $\omega = 1$, $\gamma = 0.05$, the initial state $x_0 = (0.5, 0, 0)$ and the final time $T=10$. Consider piecewise constant controls $v,n$ in the class $\mathcal{U}([0,10],Q^4)$,
i.e. with $d^q = 0.4$ that means $v_j \in [-40, 40]$, $n_j \in [0, 8]$, 
$j = \overline{0,N_v-1}$, $N_v=N_n=10$, when $v_{\min} = -100$, $v_{\max} = 100$.  

Firstly, consider the pointwise estimation $\widehat{\mathcal{R}}(10, (0.5,0,0), \mathcal{U}([0,10], Q^4))$, which was found without any regularizer and was described in Subsection~\ref{subsection7.1}, see Fig.~\ref{Fig3}(g). The estimated volume of the corresponding RS is equal to 25.4~\% of the Bloch ball's volume and is indicated in Fig.~\ref{Fig2}(a). Here the nodes $\{x^s\}$ of the grid $G(M)$, which relate to the estimation $\widehat{\mathcal{R}}(10, (0.5,0,0), \mathcal{U}([0,10], Q^4))$, are of interest for further sifting under the usage the regularizer~(\ref{f_8}). We considered ${\rm card}(\widehat{\mathcal{R}}(10, (0.5,0,0), \mathcal{U}([0,10], Q^4)))$ minimization problems of the type~(\ref{f_19}) and the corresponding OCPs of the type~(\ref{f_22}). The threshold $\delta_{x_T} = 
0.05$ was set. The following cases of the weight coefficients were taken: 

\begin{eqnarray*}
	\left(\left( \beta_{x_T}, \beta^{\rm Var}_{dv}, \beta^{\rm Var}_{dn}\right)_j \right)_{j=1}^6 &=& 
	\big( (1, ~ 5 \cdot 10^{-5}, ~ 5 \cdot 10^{-4}), ~
	(1, ~ 10^{-4}, 10^{-3}), (1, ~ 5 \cdot 10^{-4}, ~ 5 \cdot 10^{-3}), \nonumber \\ 
	&&(1, ~ 10^{-3}, ~ 10^{-2}), ~ (1, ~ 5 \cdot 10^{-3}, ~ 5 \cdot 10^{-2}), ~ 
	(1, ~ 10^{-2}, ~ 0.1) \big). 
\end{eqnarray*}
Here we worked in the frames of Algorithm~\ref{algorithm1} with DEM and DAM. For the same OCP, two attempts of DEM and two attempts of DAM were made for better guarantee. As the result, we observed how the number of such nodes, which were mentioned as reachable, depends on the indexes of these six triples. In other words, we found how the estimated volumes decrease as the index~$j$ increases: 
\[
(j)_{j=1}^6 \mapsto \left(25.4,~25.4,~25.3,~24.7,~9.1,~2 \right), \quad 
\text{in \% of the Bloch ball's volume}.
\]
Thus, for the first four triples, the estimated volumes are equal or near 
the estimated volume corresponding the case illustrated in Fig.~\ref{Fig3}(g) and found without any regularizer. For the last two triples, the estimated volumes are essentially different.

{\bf Estimating RSs of the system (\ref{f_16}) in the situation when the regularizer (\ref{f_14}) is used.}
As before, here we used $\kappa = 0.01$, $\omega = 1$, $\gamma = 0.05$, $d^q = 0.4$, $v_{\min} = -100$, $v_{\max} = 100$, $T = 10$, $\delta_{x_T} = 0.05$, 
$N_v=N_n=10$, $x_0 = (0.5, 0, 0)$. The value $\delta_{x_T} = 0.05$ gives $M(x(T), x^s, \delta_{x_T}) \leq 2 - \delta_{x_T} = 1.95$. Different results about reachability were obtained by considering different pairs $\left(\delta_{dv}, \delta_{dn} \right)$ of the thresholds in~(\ref{f_11}). In the composite objective functional~(\ref{f_22}), its weight coefficient were taken according to Statement~\ref{statement1} as shown in Table~\ref{table:1}.
\begin{table}[h!]
	\renewcommand\arraystretch{1.3} 
	\begin{tabular}{|c | c | c| c | c| c| c |}
		\hline
		~$\delta_{dv}$~  &  ~$\delta_{dn}$~  & ~ $\beta_{x_T}^{\max}$ ~ & ~$\beta_{dv}^{\max}$~ & ~$\beta_{dn}^{\max}$~ & ~${\rm card}(\widehat{\mathcal{R}})$ (\% of ${\rm card}(G(M))$)~ & ~ Estimated vol., \% of $4\pi/3$~ \\ \hline
		~ 10 ~ & ~ 0.5 ~  & ~ 36 ~  & ~1~ & ~9~ & ~ 393 ~($\approx 9.4$ \%)~ & ~$\approx 9.4$ ~  \\ \hline
		~ 20 ~ & ~ 1 ~  & ~ 31 ~  & ~1~ & ~9~ & ~749~($\approx 18$ \%)~ & ~$\approx 17.9$ ~ \\ \hline
		~ 40 ~ & ~ 2 ~  & ~ 21 ~ & ~1~ & ~7~ & ~1041~($\approx 25$ \%)~ & ~$\approx 24.9$ ~\\ \hline	
	\end{tabular}  
	\caption{Computing the RSs' pointwise estimations for $x_0 = (0.5,0,0)$ and $T=10$.}
	\label{table:1}
\end{table}

As the first example, we considered $\delta_{dv} = 10$ and  $\delta_{dn} = 0.5$. For this case, the formulas~(\ref{f_48}), (\ref{f_49}) give the following inequalities: $M^{\delta_{dv}} \leq 70$ and  $M^{\delta_{dn}} \leq 7.5$. Compare them with each other and with the inequality $M(x(T), x^s, \delta_{x_T}) \leq 2 - \delta_{x_T} = 1.95$, the weight coefficients $\beta_{dv}^{\max} = 1$, $\beta_{x_T}^{\max} = \left[ 70/1.95 \right] = 36$, and  $\beta_{dn}^{\max} = \left[ 70/7.5 \right] = 9$ were taken for balancing the terms in the objective functional in (\ref{f_22}). Here  
DEM and DAM were used.  As the result, reachability only of 393 nodes of the grid $G(M)$ was established, i.e. near 9.4~\% of ${\rm card}(G(M))$; the estimated volume of the RS is near 9.4~\% of the Bloch ball's volume equal to $4\pi/3$.

In the second example, we set $\delta_{dv} = 20$ and $\delta_{dn} = 1$. For this case, the formulas (\ref{f_48}), (\ref{f_49}) give $M^{\delta_{dv}} \leq 60$ and  $M^{\delta_{dn}} \leq 7$. Then the weight coefficients $\beta_{dv}^{\max} = 1$, $\beta_{x_T}^{\max} = \left[ 60/1.95 \right] = 31$, and $\beta_{dn}^{\max} = \left[ 60/7 \right] = 9$ were taken. Here reachability of 749 nodes of the grid $G(M)$ was established, i.e. near 18~\% of ${\rm card}(G(M))$.

In the third example, we used $\delta_{dv} = 40$ and $\delta_{dn} = 2$. For this case, the formulas (\ref{f_48}), (\ref{f_49}) give $M^{\delta_{dv}} \leq 40$ and  $M^{\delta_{dn}} \leq 6$. Here the weight coefficients $\beta_{dv}^{\max} = 1$, $\beta_{x_T}^{\max} = \left[ 40/1.95 \right] = 21$, and $\beta_{dn}^{\max} = \left[ 40/6 \right] = 7$ were set. Here reachability of 1041 nodes was found, i.e. near 25~\%  of ${\rm card}(G(M))$.

These numerical results are also given in Table~\ref{table:1}. We see that decreasing $\delta_{dv}$, $\delta_{dn}$ leads to decreasing numbers of reachable nodes, i.e. it gives the situation, when for some part of nodes, which were found as reachable for larger values of these thresholds, the algorithm did not find admissible controls $v,n$, which could transfer the system from the given $x_0$ to these nodes. 

{\bf Estimating CSs of the system (\ref{f_16}) in the situation when the regularizer (\ref{f_14}) is used.} 
Here the initial state $x_0$ is not fixed. We set the target state $x_{\rm target} = (0.5, 0, 0)$. As before, we used $\kappa = 0.01$, $\omega 
= 1$, $\gamma = 0.05$, $d^q = 0.4$, $v_{\min} = -100$, $v_{\max} = 100$, $T = 10$, $\delta_{x_T} = 0.05$, $N_v=N_n=10$. We worked in the frames of Algorithm~\ref{algorithm2} here. The weight coefficients of the objective functional (\ref{f_28}) were set also with usage of the inequalities (\ref{f_48}), (\ref{f_49}), and $M(x(T), x_{\rm target}, \delta_{x_T}) \leq 2 - \delta_{x_T} = 1.95$ for $\delta_{x_T} = 0.05$. The corresponding information is given in Table~\ref{table:2}.
\begin{table}[h!]
	\renewcommand\arraystretch{1.3}   
	\begin{tabular}{|c | c | c| c | c| c| c| c |}
		\hline
		~$\delta_{dv}$~ &  ~$\delta_{dn}$~ & ~ $\beta_{x_T}^{\max}$ ~ & ~$\beta_{dv}^{\max}$~ & ~$\beta_{dn}^{\max}$~ & ~${\rm card}(\widehat{\mathcal{C}})$ (\% of ${\rm card}(G(M))$) ~ & ~ Estimated vol., \% of $4\pi/3$~ \\ \hline
		~ 20 ~ & ~ 1 ~ & ~ 31 ~ & ~1~ & ~ 9 ~ & ~ 3228~($\approx 77.4$ \%) ~ & ~ $\approx 77.1$ ~  \\ \hline
		~ 40 ~ & ~ 2 ~ & ~ 21 ~ & ~1~ & ~ 7 ~ & ~ 4093 ($\approx 98.2$ \%) ~ & ~ $\approx 97.7$ ~ \\ \hline
	\end{tabular} \\
	\caption{Computing the CSs' pointwise estimations with $x_{\rm target} = 
(0.5,0,0)$ and $T=10$.}
	\label{table:2}
\end{table} 

As the first example, we consider $\delta_{dv} = 20$ and  $\delta_{dn} = 1$. As before, the values $\beta_{dv}^{\max} = 1$, $\beta_{x_T}^{\max} = 31$, and $\beta_{dn}^{\max} = 9$ were taken. The computed pointwise estimation of the CS consists of 3228 nodes, i.e. 77.4~\% of the cardinality of $G(M)$.

In the second example, we set $\delta_{dv} = 40$ and $\delta_{dn} = 2$. The values $\beta_{dv}^{\max} = 1$, $\beta_{x_T}^{\max} = 21$, and 
$\beta_{dn}^{\max} = 7$ were taken. The obtained pointwise estimation of the CS  consists of 4093 nodes, i.e. 98.2~\% of the cardinality of $G(M)$.

The described above results (see Table~\ref{table:2}) show that increasing the thresholds $\delta_{dv}$, $\delta_{dn}$ increases the number of nodes, from which the system is moved to the given target state $x_{\rm target}$. Comparing the results shown in Tables~\ref{table:1},~\ref{table:2}, 
we see that the number of the nodes $\{x^s \} \subset G(M)$, from which the system is moved to the given target state $x_{\rm target} = (0.5,0,0)$, is essentially larger than the number of the nodes, to which the system is moved from the initial state $x_0 = (0.5, 0, 0)$, for the same conditions ($T = 10$, etc.).  


\section{\label{Conclusions}Conclusions}

In this article, an open two-level quantum system~\cite{Pechen2011,Morzhin_Pechen_IJTP_2019, Morzhin_Pechen_LJM_2019, Morzhin_Pechen_Physics_of_Particles_and_Nuclei, Morzhin_Pechen_LJM_2020, Morzhin_Pechen_SteklovProceedings},  
whose evolution is governed by the Gorini--Kossakowski--Lindblad--Sudarshan master equation with Hamiltonian and dissipation superoperator depending, correspondingly,
on coherent and incoherent controls, was considered. Using the Bloch parametrization, which gives
bijection between density matrices and 3-dimensional real vectors, we analyzed  in terms of Bloch vectors 
the corresponding dynamical system and the problem of estimating RSs and CSs. In addition to the constraint on controls' magnitudes, different types for constraining controls' variations were written and taken into account in the definitions of a RS and a CS of the system in the terms of Bloch vectors. In the article, the idea of estimating RSs by considering their sections~\cite{Morzhin_Tyatyushkin_2008, Tyatyushkin_Morzhin_2011} was 
used in the definitions of pointwise estimations of RSs and CSs, and also 
in the corresponding algorithms. These algorithms are based on solving series of OCPs being here finite-dimensional optimization problems, because 
piecewise constant controls are considered. For solving these optimization problems, DEM and DAM were applied, at that, for each optimization problem, several runs of DEM and/ or DAM were done excepting a case, when an objective function a priori is non-negative and the first run of DEM or DAM gives zero value for this objective function. 

For some specific values of the system's parameters $\omega$, $\mu$, $\gamma$, the bounds $v_{\min}$, $v_{\max}$, $n_{\max}$, the thresholds $\delta_{dv}$, $\delta_{dn}$, the computational experiments were performed. The numerical results, which are described in Section~\ref{section7}, show how the RSs' estimations depend on distances between the system's initial 
states and the Bloch ball's center point, final times, constraints on controls' magnitudes and variations. Subsection~\ref{subsection7.2} shows how the cardinalities of the RSs' and CSs' pointwise estimations and the  estimated volumes depend on changing the weight coefficients and the thresholds in the corresponding objective functionals, which contain the regularizers for additional constraining controls' variations. The numerical results described in Section~\ref{section7} show that: (a)~additional constraints on controls can essentially decrease the estimated volumes of RSs 
(see Fig.~\ref{Fig2}) and CSs, i.e., in other words, control possibilities to steer the system from one state to another state over some time range; (b)~changing the final time $T$ also can essentially decrease volumes and geometry of RSs (see Fig.~\ref{Fig2}--\ref{Fig5}); (c)~estimated volumes of RSs can essentially depend on selecting the initial state~$x_0$ (compare Fig.~\ref{Fig4} and Fig.~\ref{Fig5}, where $x_0$ represents, correspondingly, either the completely mixed or some pure quantum state); (d)~it can be reasonable to look for some trade-off between, on one hand, control possibilities to steer the system from one state to another state and, on other hand, looking for more appropriate control probably in the terms of decreasing the final time and controls' variations.  
 
\vspace{0.4cm}

\leftline{\bf \Large Acknowledgments}

\vspace{0.25cm}

This article was performed in Steklov Mathematical
Institute of Russian Academy of Sciences within the project of the Russian Science Foundation No.~17-11-01388.

\end{document}